\documentclass[prl,twocolumn,superscriptaddress,longbibliography]{revtex4-2}
\usepackage{amsmath}
\usepackage{amssymb}
\usepackage{amsthm}
\usepackage{amsfonts}
\usepackage[version=4]{mhchem}
\usepackage{listings}
\usepackage{enumerate}
\usepackage{latexsym}
\usepackage{psfrag}
\usepackage{bm}
\usepackage[all]{xy}
\usepackage{subfigure}
\usepackage[pdftex,colorlinks]{hyperref}
\usepackage{color}
\usepackage{mathtools}
\usepackage{times}
\usepackage{array}
\usepackage{placeins}
\usepackage{siunitx}
\begin{document}

\title{Berezinskii-Kosterlitz-Thouless region and magnetization plateaus in easy-axis triangular weak-dimer antiferromagnet \ce{K2Co2(SeO3)3}}

\author{Ying Fu}
\affiliation{Quantum Science Center of Guangdong-Hong Kong-Macao Greater Bay Area (Guangdong), Shenzhen 518045, China}
\affiliation{Department of Physics, Southern University of Science and Technology, Shenzhen 518055, China} 

\author{Han Ge}
\author{Jian Chen}
\author{Jie Xiao}
\author{Yi Tan}
\affiliation{Department of Physics, Southern University of Science and Technology, Shenzhen 518055, China}
\author{Le Wang}
\affiliation{International Quantum Academy, Shenzhen, 518048, China}

\author{Junfeng Wang}
\author{Chao Dong}
\affiliation{Wuhan National High Magnetic Field Center, Huazhong University of Science and Technology, Wuhan 430074, China}

\author{Zhe Qu}
\author{Miao He}
\affiliation{Science Island Branch of Graduate School, University of Science and Technology of China, Hefei, Anhui, 230026, China}
\affiliation{Anhui Key Laboratory of Low-Energy Quantum Materials and Devices, High Magnetic Field Laboratory, HFIPS, Chinese Academy of Sciences, Hefei, Anhui 230031, China}

\author{Chuanying Xi}
\author{Langsheng Ling}
\affiliation{Anhui Key Laboratory of Low-Energy Quantum Materials and Devices, High Magnetic Field Laboratory, HFIPS, Chinese Academy of Sciences, Hefei, Anhui 230031, China}
\author{Bin Xi}
\email{xibin@yzu.edu.cn}
\affiliation{College of Physics Science and Technology, Yangzhou University, Yangzhou 225002, China}

\author{Jia-Wei Mei}
\email{meijw@sustech.edu.cn}
\affiliation{Department of Physics, Southern University of Science and Technology, Shenzhen 518055, China} 
\date{\today}

\begin{abstract}	
We investigate the magnetic phase diagram of the bilayer triangular antiferromagnet \ce{K2Co2(SeO3)3}, revealing a rich interplay among geometric frustration, bilayer coupling, and symmetry-driven phenomena. High-field magnetization measurements show fractional magnetization plateaus at 1/3, 1/2, 2/3, and 5/6 of the saturation magnetization. To elucidate the experimental magnetic phase diagram at low fields, we propose that \ce{K2Co2(SeO3)3} can be described as an easy-axis triangular weak-dimer antiferromagnet. We emphasize the critical role of the emergent \(U(1) \otimes S_3\) symmetry, where \(S_3 = \mathbb{Z}_3 \otimes \mathbb{Z}_2^d\), in determining the magnetic phases at low fields. The remarkable agreement between the experimental and theoretical phase diagrams suggests that the phase transitions are governed by this symmetry. Notably, our combined experimental and theoretical results identify a Berezinskii-Kosterlitz-Thouless (BKT) phase region at finite fields. These findings provide new insights into the phase structure of frustrated magnets and establish \ce{K2Co2(SeO3)3} as a compelling platform for exploring unconventional quantum phenomena in \(U(1) \otimes S_3\) systems.
\end{abstract}
\maketitle
\emph{Introduction. --}
The triangular-lattice antiferromagnet serves as a paradigmatic platform for investigating the interplay of geometric frustration and quantum fluctuations~\cite{Zapf2014}. The inherent geometric frustration in triangular lattices prevents all spins from aligning antiferromagnetically, leading to a highly degenerate ground state and amplifying quantum fluctuations~\cite{Diep2005}. This unique interplay has been extensively studied in various $3d$ transition-metal compounds, including \ce{Cs2CuBr4}~\cite{Ono2003,Alicea2009}, \ce{Ba3CoSb2O9}~\cite{Zhou2012,Susuki2013}, \ce{Ba3MnNb2O9}~\cite{Lee2014}, \ce{Na2BaCo(PO4)2}~\cite{Zhong2019,Li2020,Sheng2022}, \ce{Na2BaNi(PO4)2}~\cite{Li2021,Sheng2025}, \ce{K2Co(SeO3)2}~\cite{Zhong2020a,Zhu2024,Chen2024a}, and \ce{Rb4Mn(MoO4)3}~\cite{Ishii2009}. Recent research has also highlighted the significance of bilayer triangular lattices, such as $A_3$\ce{Cr2O8} ($A$ = Sr, Ba)~\cite{Aczel2009,Aczel2009a,Kofu2009}, $A_2$\ce{Co2(SeO3)3} ($A$ = K, Rb)~\cite{Zhong2020,Xu2024}, and \ce{K2Ni2(SeO3)3}~\cite{Li2024,Yue2024}, which offer additional opportunities to explore the effects of interlayer interactions and field-induced phase transitions~\cite{Yamamoto2013,Strecka2018,Chen2024}.

\begin{figure}[b]
	\centering
	\includegraphics[width=1\columnwidth]{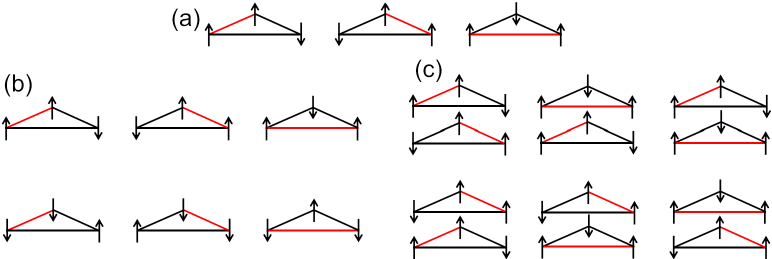}
	\caption{
		(a) Spin configurations illustrating \(\mathbb{Z}_3\) symmetry in an antiferromagnetic triangle. (b) Spin configurations demonstrating \(\mathbb{Z}_6\) symmetry in a single-layer triangle with time-reversal symmetry. (c) Spin configurations corresponding to \(S_3 = \mathbb{Z}_3 \otimes \mathbb{Z}_2^d\) symmetry in a weak-dimer bilayer triangle system.}
	\label{fig:fig1}
\end{figure}

One of the most striking features of triangular lattice systems is the 1/3 magnetization plateau, which is due to the order-by-disorder effect~\cite{Chubukov1991,Sheng1992}, characterized by the collinear up-up-down (\(uud\)) spin configuration. This phase exemplifies a spontaneous breaking of the threefold sublattice permutation symmetry \(\mathbb{Z}_3\) (Fig.~\ref{fig:fig1}(a))~\cite{Miyashita1986,MIYASHITA2010,Melchy2009}, which falls into the universality class of the three-state clock model~\cite{Jose1977,Wu1982}. Quantum fluctuations further enhance the stability of this plateau phase~\cite{Chubukov1991,Sheng1992,Alicea2009}. At zero magnetic field, the time-reversal symmetry enlarges \(\mathbb{Z}_3\) to \(\mathbb{Z}_6\)~\cite{Melchy2009} (Fig.~\ref{fig:fig1}(b)), which stabilizes an intermediate Berezinskii-Kosterlitz-Thouless (BKT) phase with quasi-long-range spin correlations with algebraic decay $\langle S(0)S(r)\rangle\sim r^{-\eta}$~\cite{Jose1977,Nienhuis1984,Melchy2009}, where the anomalous dimension $\eta$ quantifies deviation from the classical mean-field dimension. The symmetry connection between the 1/3 plateau and the intermediate BKT phase highlights the critical role of symmetry in shaping phase transitions~\cite{Landau1937,Ji2020,Chatterjee2023a}. However, the intermediate BKT phase is fragile in the presence of a magnetic field~\cite{Hu2020}, and its experimental observation remains elusive.

In this work, we propose that the bilayer triangular antiferromagnet \ce{K2Co2(SeO3)3} exemplifies the weak-dimer regime in bilayer systems. Experimentally, we observe successive magnetic phase transitions under low magnetic fields, accompanied by a series of magnetization plateaus at 1/3, 1/2, 2/3, and 5/6 of the saturation magnetization. Theoretically, the bilayer dimer symmetry (\(\mathbb{Z}_2^d\)) enriches the triangular lattice's \(\mathbb{Z}_3\) symmetry, extending it to \(S_3 = \mathbb{Z}_3 \otimes \mathbb{Z}_2^d\) (Fig.~\ref{fig:fig1}(c)), while preserving a continuous \(U(1)\) symmetry associated with \(c\)-axis spin rotations. By combining experimental measurements and theoretical modeling, we construct the magnetic phase diagram with a particular focus on the low-field regime. In addition to the ``Y'' phase, associated with \(U(1)\) symmetry breaking of the \(c\)-axis spin rotation symmetry, our study reveals three distinct phases driven by the breaking of the discrete \(S_3\) symmetry: the \(S_3\)-broken (triplet-singlet-singlet, $tss$) phase, the \(\mathbb{Z}_3\)-broken (\(uud\)) phase, and an intermediate BKT phase. These findings depart from prior investigations on strong-dimer systems~\cite{Aczel2009,Kofu2009,Yamamoto2013,Strecka2018}. Notably, the theoretical phase diagram aligns closely with experimental observations, underscoring the critical role of symmetry in governing the phase transitions.

\emph{Methods. --}
We synthesized \ce{K2Co2(SeO3)3} single crystals using the flux method, as detailed in previous reports \cite{Zhong2020}. Magnetization, heat capacity, and magnetocaloric effect (MCE) measurements were conducted using the Quantum Design Magnetic Property Measurement System (up to 7 T) and the Physical Property Measurement System (up to 14 T). High-field magnetization measurements up to 35 T were carried out using a Vibrating Sample Magnetometer (VSM) on a water-cooled magnet (WM5) at the High Magnetic Field Laboratory (HMFL). The Monte Carlo (MC) simulations were executed employing a combination of the parallel tempering method~\cite{Hukushima1996} and the heat-bath algorithm~\cite{Miyatake1986}.

\emph{Weak-dimer interaction and multiple magnetization plateaus in \ce{K2Co2(SeO3)3}. --}
\begin{figure}[b]
	\centering
	\includegraphics[width=\columnwidth]{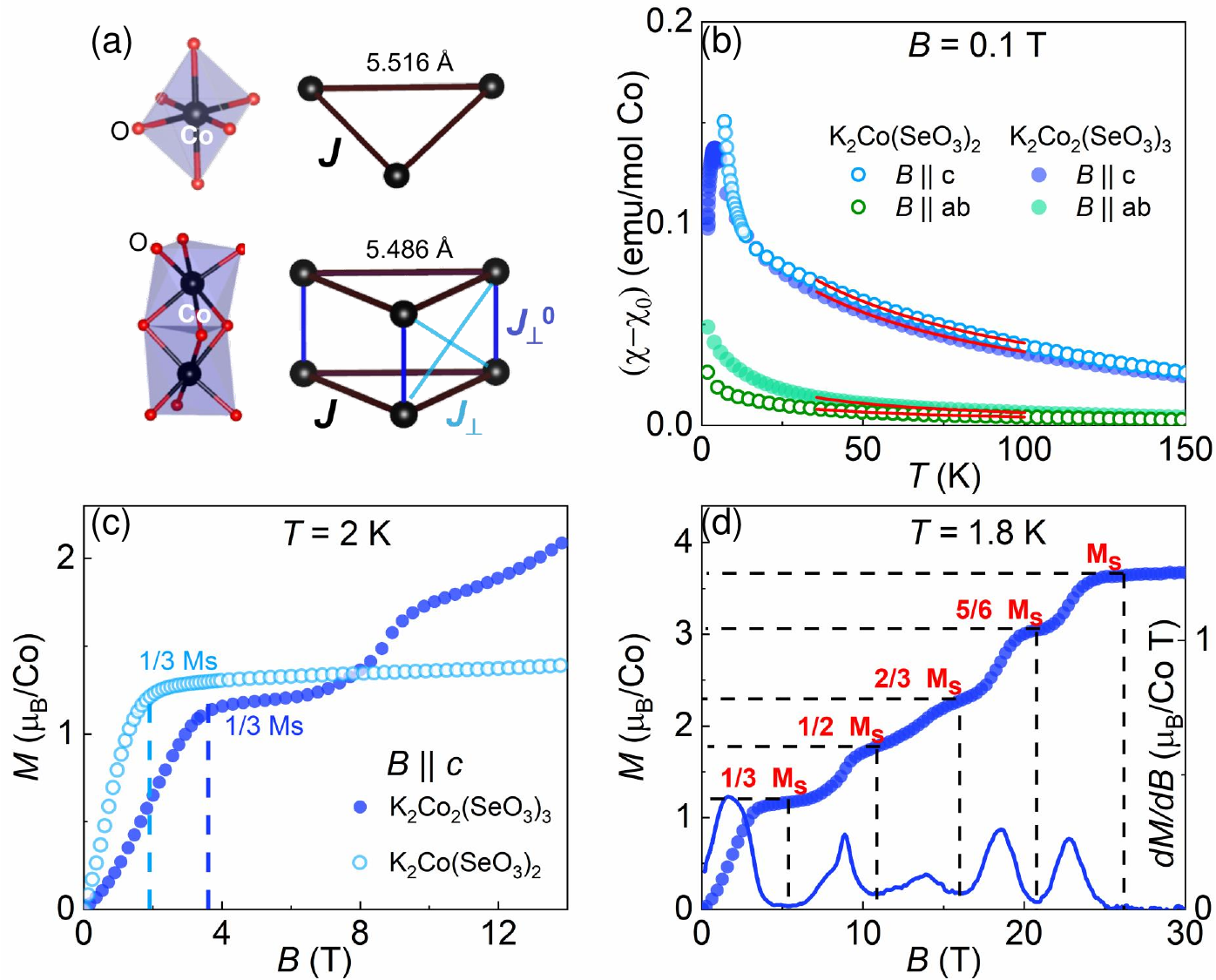}
	\caption{(a) CoO$_6$ and Co$_2$O$_9$ polyhedra forming the monolayer and bilayer triangular antiferromagnets of \ce{K2Co(SeO3)2} and \ce{K2Co2(SeO3)3}, respectively. (b) Magnetic susceptibilities \( \chi = M/H \) for \ce{K2Co(SeO3)2} and \ce{K2Co2(SeO3)3}, with the temperature-independent \( \chi_0 \) subtracted based on Curie-Weiss fits (red lines). (c) Field-dependent magnetization curves for \ce{K2Co(SeO3)2} and \ce{K2Co2(SeO3)3} at 2~K. (d) High-field magnetization and its derivative for \ce{K2Co2(SeO3)3} with $B \parallel c$ at 1.8~K.}
	\label{fig:Fig2}
\end{figure}
\begin{figure*}[t]
	\centering
	\includegraphics[width=2\columnwidth]{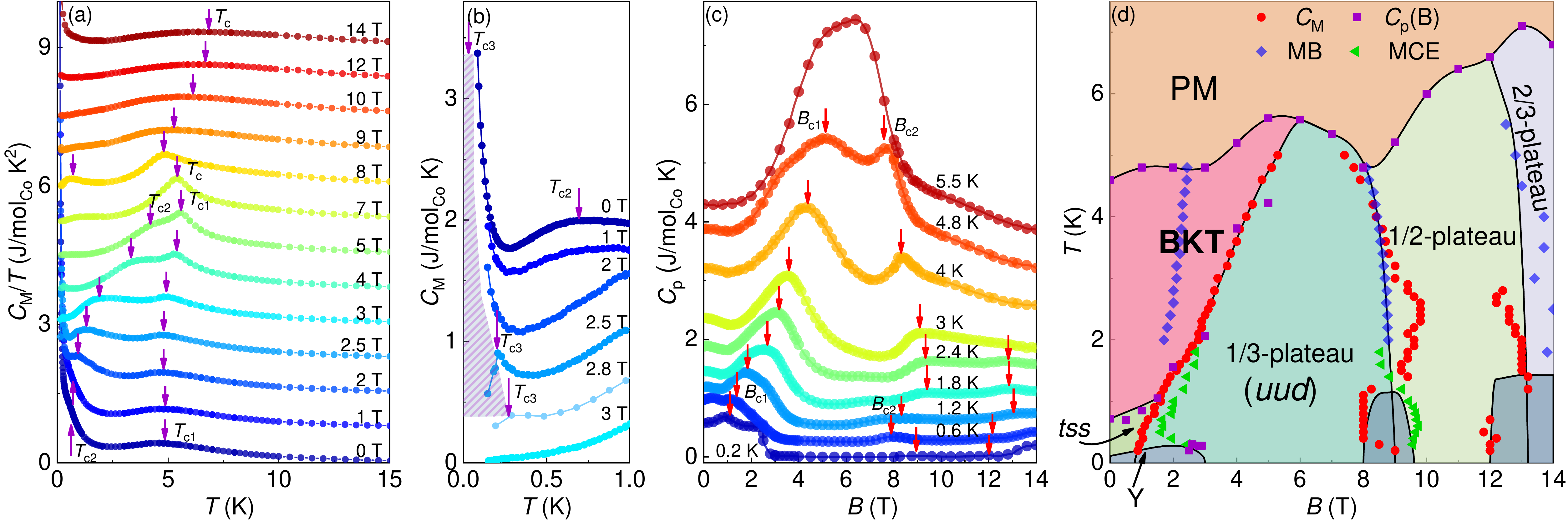}
	\caption{(a) Magnetic specific heat divided by temperature $C_m(T)/T$ (offset of 0.75) of \ce{K2Co2(SeO3)3} for $B \parallel c$ axis. (b) Zoomed-in view of $C_m(T)$ (offset of 0.25) in (a) to resolve the phase transition at low field and temperature. (c) Field-dependent heat capacity $C_p$ (offset of 0.25). (d) Magnetic phase diagram of \ce{K2Co2(SeO3)3} for $B \parallel c$ axis. The lines for the phase boundaries are guides for the eye. The transition points are extracted from thermodynamic measurements. Phase labels for BKT, $tss$, $uud$ and ``Y'' are based on subsequent theoretical simulations.}
	\label{fig:Fig3}
\end{figure*}
As shown in Fig.~\ref{fig:Fig2}(a), along with its monolayer counterpart \ce{K2Co(SeO3)2}~\cite{Zhong2020a,Zhu2024,Chen2024a}, the bilayer triangular antiferromagnet \ce{K2Co2(SeO3)3} has attracted considerable attention~\cite{Zhong2020,Chen2024}. The dominant intralayer interaction $J$ is present in both \ce{K2Co2(SeO3)3} and \ce{K2Co(SeO3)2}. $J_\bot^0$ is the intra-dimer interaction, and $J_\bot$ is the crossed-layer interaction.

The intra-dimer interaction $J_\bot^0$ in \ce{K2Co2(SeO3)3} arises from a combination of ferromagnetic superexchange via shared oxygen atoms, as per the Goodenough-Kanamori rules~\cite{Goodenough2008} for an 85\textdegree~ Co-O-Co bond angle, and antiferromagnetic direct exchange between \ce{Co^{2+}} ions. Unlike systems such as \ce{Cs3V2Cl9}~\cite{Leuenberger1986} and \ce{Cs3Fe2Cl9}~\cite{Ishii2021,Gao2024}, where ferromagnetic superexchange is predominant, the direct exchange interaction in \ce{K2Co2(SeO3)3} plays a significant role. Contrary to the strong dimer case~\cite{Aczel2009,Kofu2009,Yamamoto2013,Strecka2018}, we suggest that \ce{K2Co2(SeO3)3} is more accurately described as a weak-dimer bilayer triangular antiferromagnet, due to the competing exchanges, despite the short Co-Co bonds within the \ce{Co2O9} dimers indicating dominant intra-dimer interactions.

The weak intra-dimer interaction in \ce{K2Co2(SeO3)3} is evidenced by the comparative study of magnetic susceptibilities for the two compounds, shown in Fig.~\ref{fig:Fig2}(b). In the relevant temperature range \(40~\mathrm{K} < T < 100~\mathrm{K}\), where the effective spin-\(1/2\) state of \ce{Co^{2+}} ions emerges, both \ce{K2Co2(SeO3)3} and \ce{K2Co(SeO3)2} exhibit paramagnetic behavior. Their effective Curie-Weiss temperatures can be used to estimate the exchange interactions, \(\Theta_{\mathrm{CW}} = 1.5(J + J_\perp + J_\perp^0/6)\) for \ce{K2Co2(SeO3)3} and \(\Theta_{\mathrm{CW}} = 1.5J\) for \ce{K2Co(SeO3)2}. As indicated in Fig.~\ref{fig:Fig2}(b), both compounds share comparable magnetic characteristics. With \(\Theta_{\mathrm{CW}}^c = -43~\mathrm{K}\) for \ce{K2Co2(SeO3)3} and \(-49~\mathrm{K}\) for \ce{K2Co(SeO3)2}, these values are remarkably close, suggesting an effectively weak-dimer interaction (\(J_\perp^{\mathrm{eff}} = J_\perp + J_\perp^0/6\)).

Figure~\ref{fig:Fig2}(c) compares the magnetization curves of \ce{K2Co2(SeO3)3} and \ce{K2Co(SeO3)2} at 2~K. Despite the large Curie-Weiss temperatures (exceeding 40~K), the critical fields for the onset of the 1/3 magnetization plateau are relatively small, a behavior attributed to the strong magnetic anisotropy in both systems. The close proximity of the critical fields, with their difference being significantly smaller than the Curie-Weiss temperature scale, further supports the classification of \ce{K2Co2(SeO3)3} as a weak-dimer system. Notably, the absence of an obvious spin gap at 2 K near zero field in the magnetization of \ce{K2Co2(SeO3)3} underscores its weak intra-dimer interaction, contrasting sharply with the pronounced spin gaps observed in field-dependent magnetization curves of strong-dimer systems such as~\ce{Sr3Cr2O8}\cite{Aczel2009a} and \ce{SrCu2(BO3)2}~\cite{Onizuka2000}.

High-field magnetization measurements along the \(c\)-axis at 1.8 K, shown in Fig.~\ref{fig:Fig2}(d) and Fig.S1 in Supplementary Materials~\cite{notes}, reveal a series of distinct magnetization plateaus. The derivative curve (blue line) highlights the precise locations of these plateaus, which occur at fractional values of the saturation magnetization: 1/3, 1/2, 2/3, and 5/6. The compound reaches saturation at approximately \(25~\mathrm{T}\) with a magnetic moment of \(\sim 3.7~\mu_B\), closely matching the saturation field (\(\sim 22~\mathrm{T}\)) and moment (\(\sim 3.9~\mu_B\)) of its monolayer counterpart \ce{K2Co(SeO3)2}~\cite{Zhu2024}. 
Despite being in the weak-dimer regime, bilayer-dimers in \ce{K2Co2(SeO3)3} amplify geometric frustration (Fig.~\ref{fig:fig1}), likely driving the emergence of multiple magnetization plateaus.

\emph{Phase diagram below 14~T in \ce{K2Co2(SeO3)3}. -- }
Figure~\ref{fig:Fig3}(a) presents the magnetic specific heat $C_m(T)$ with the phonon contribution subtracted (see Fig.S3 in Supplementary Materials~\cite{notes}), as a function of temperature for magnetic fields applied along the \( c \)-axis of \ce{K2Co2(SeO3)3}. Given the effective interaction around 40~K, as indicated by the Curie-Weiss fitting, the peak features observed between 5~K and 10~K likely represent phase transitions rather than mere correlations. The broad nature of these transitions reflects the two-dimensional characteristics of the system.

To better resolve these transitions, we first examine the specific heat at \( B = 7~\mathrm{T} \) (see Fig.~\ref{fig:Fig3}(a)), where the sharpest feature is observed, indicating a single phase transition at \( T_c = 5.4~\mathrm{K} \). For fields above \( 7~\mathrm{T} \) and up to \( 14~\mathrm{T} \), only one phase transition persists, though it becomes increasingly broad and challenging to pinpoint precisely. Below \( 7~\mathrm{T} \), the specific heat peak splits into two, indicating the presence of two distinct phase transitions at \( T_{c1} \) and \( T_{c2} \). As the magnetic field decreases, the position of \( T_{c1} \) decreases slightly but broadens, while \( T_{c2} \) decreases more rapidly and remains broad. At even lower fields, below \( 3~\mathrm{T} \), an additional phase transition $T_{c3}$ emerges around 0.2~K. To provide a clearer view of this low-temperature behavior, Fig.~\ref{fig:Fig3}(b) zooms in on the specific heat data below \( 1.0~\mathrm{K} \), highlighting the transition at \( T_{c3} \). 

Figure~\ref{fig:Fig3}(c) presents field-dependent specific heat $C_p(B)$ measurements for \ce{K2Co2(SeO3)3} at various fixed temperatures. For instance, at \( T = 0.6~\mathrm{K} \), the specific heat curves exhibit two concave regions in the ranges of 3–8~T and 9–13~T, corresponding to the 1/3 and 1/2 plateaus (Fig.~\ref{fig:Fig2}(d)), respectively. Around 13~T with increasing fields, the heat capacity increases and then begins to drop around 14~T (Fig.S5 in Supplementary Materials\cite{notes}), signaling the approach to the 2/3 plateau, corroborated by the high-field magnetization curve in Fig.~\ref{fig:Fig2}(d). At lower temperatures, additional features near these transitions reveal the presence of extra phases between the 1/3- and 1/2-plateau phases and between the 1/2- and 2/3-plateau phases.

Focusing on the low-field region, below \( 6~\mathrm{T} \), the transition from the zero-field phase to the 1/3-plateau phase is observed at the critical field \( B_c(T) \). For \( T > 1~\mathrm{K} \), \( B_c(T) \) above 1.5~T in Fig.~\ref{fig:Fig3}(c) aligns well with the phase boundary \( T_{c2}(B) \) between 1 K and 4.2 K, identified in Fig.~\ref{fig:Fig3}(a), validating the consistency of temperature- and field-dependent results. In contrast, for \( T < 1~\mathrm{K} \), \( B_{c1}(T) \)  between \( 0.85~\mathrm{T} \) and \( 1.5~\mathrm{T} \) deviates from \( T_{c2}(B) \) below 1 K. This discrepancy suggests the emergence of a new low-temperature phase. 

Figure~\ref{fig:Fig3}(d) presents the magnetic phase diagram of \ce{K2Co2(SeO3)3}, constructed from heat capacity, magnetization, and MCE (with detailed data in Fig.S4 in Supplementary Materials~ \cite{notes}). Besides the key features of the 1/3-, 1/2-, and 2/3-plateau phases, the low-field region (\( B < 6~\mathrm{T} \)) reveals a remarkably intricate phase structure, which is the primary focus of our study. 

Below 6~T, the system transitions from the paramagnetic phase at high temperature to the 1/3-plateau phase at low temperature through an intermediate phase. In this work, we identify this intermediate phase as BKT phase.
As the magnetic field decreases further, another transition occurs below \( 3~\mathrm{T} \) and around \( 0.1~\mathrm{K} \), leading to the emergence of the ``Y'' phase, associated with \( U(1) \) symmetry breaking. 
In the low field and temperature regime (\( B < 1~\mathrm{T}, 0.2~\rm{K}<T < 1~\mathrm{K} \)), we identify a distinct phase, termed the \( tss \) (triplet-singlet-singlet) phase, based on theoretical simulation below.

\emph{Theoretical phase diagram for \(U(1) \otimes S_3\) symmetry breaking. -- }
To  capture the essential features of experimental phase diagram in Fig.~\ref{fig:Fig3}(d), we begin with the Hamiltonian for \ce{K2Co2(SeO3)3}, given by:
\begin{eqnarray}\label{eq:ham}
	H&=&\sum_{n\langle ij\rangle}J[(S_{ni}^xS_{nj}^x+S_{ni}^yS_{nj}^y)/\Delta+ S_{ni}^zS_{nj}^z]\nonumber\\
	&+&\sum_{i}J_\bot^0[(S_{1i}^xS_{2i}^x+S_{1i}^yS_{2i}^y)/\Delta+ S_{1i}^zS_{2i}^z]\nonumber\\
	&+&\sum_{\langle ij\rangle'}J_\bot[(S_{1i}^xS_{2j}^x+S_{1i}^yS_{2j}^y)/\Delta+ S_{1i}^zS_{2j}^z]\nonumber\\ 
	&-&B\sum_{i}(S_{1i}^z+S_{2i}^z).
\end{eqnarray}
Here, \(J>0\), \(J_\bot>0\) and \(J_\bot^0>0\) are antiferromagnetic interactions as shown in Fig.~\ref{fig:Fig2}(a). The anisotropy parameter \(\Delta\) is assumed to be uniform across all bonds in the triangular plane.
\begin{figure}[t]
	\centering
	\includegraphics[width=\columnwidth]{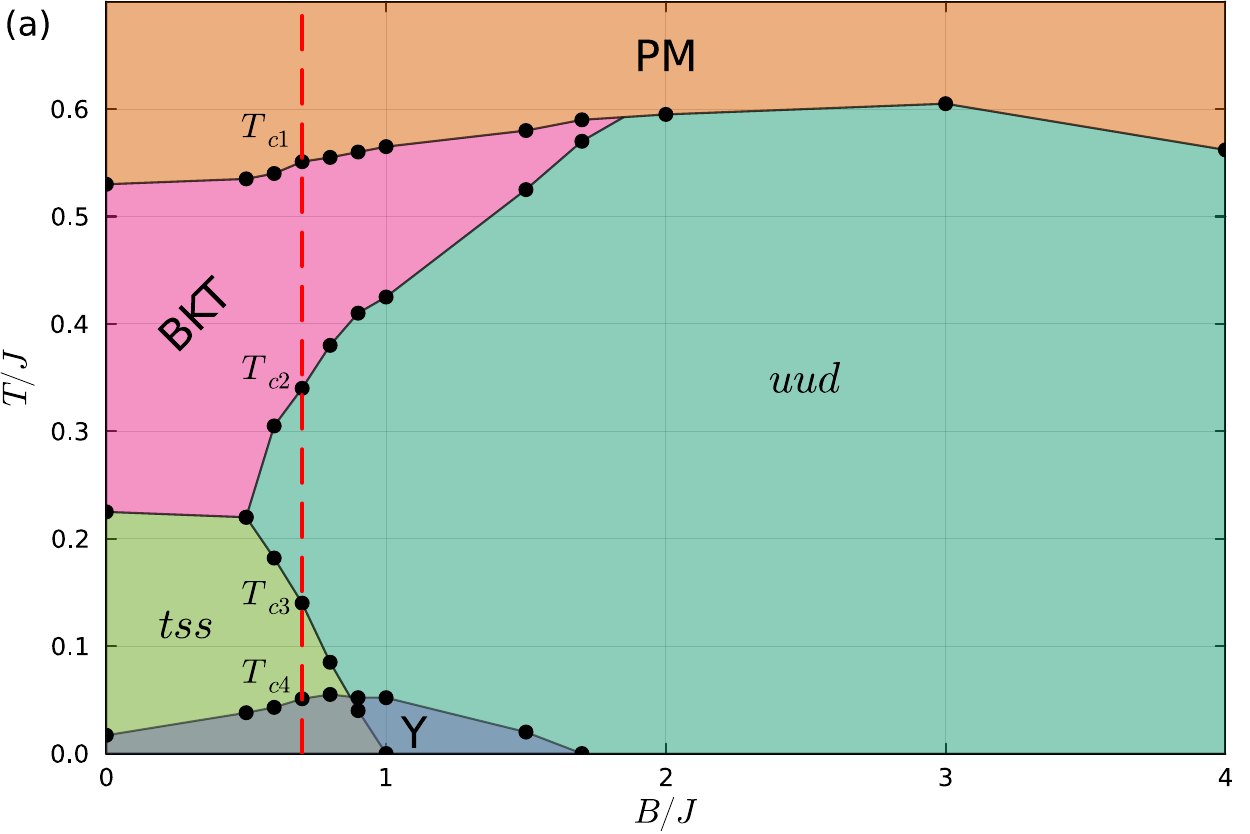}
	\includegraphics[width=\columnwidth]{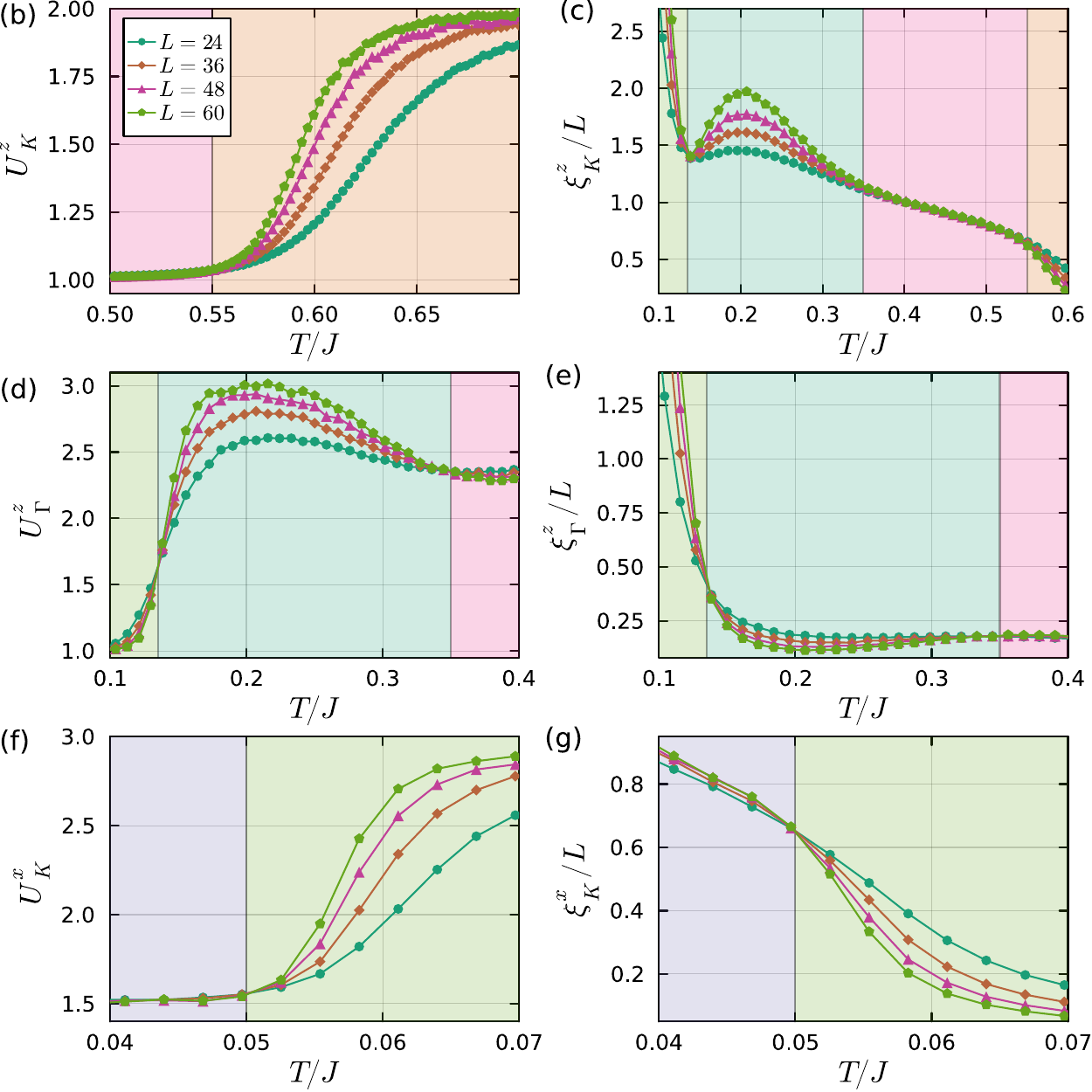}
	\caption{(a) Phase diagram obtained from classical Monte Carlo simulations for the spin dimer system described by the model (\ref{eq:ham}) with parameters \( J_\perp^0 = J \), \( J_\perp = 0.2J \), anisotropy \(\Delta = 3\), and \( J > 0 \). The paramagnetic (PM) state preserves the \( U(1) \otimes S_3 \) symmetry. The ``Y''  phase is associated with \( U(1) \) symmetry breaking. The \( uud \) state corresponds to the 1/3-magnetization plateau and is \(\mathbb{Z}_2^d\)-symmetric. The \( tss \) phase, representing the triplet-singlet-singlet configuration, corresponds to one of the six spin states shown in Fig.~\ref{fig:fig1}(c) and breaks \( S_3 \) symmetry. The BKT phase is characterized by intermediate algebraic spin correlations. (b)-(g) Binder ratio \( U \) and scaled correlation length \( \xi/L \) for various order parameters \( \psi_K^z \), \( \psi_\Gamma^z \), and \( \psi_K^x \) defined in Eq.(\ref{eq:order}) are shown for \( B/J = 0.7 \).}
	\label{fig:Fig4}
\end{figure}

To explore the phase diagram, we perform classical Monte Carlo simulations on \(L \times L \times 2\) clusters (\(L = 24, 36, 48, 60\)) with periodic boundary conditions in the plane. The order parameters are defined as
\begin{eqnarray}\label{eq:order}
	\psi_{K}^x&=&\frac{1}{L}\sum_{i}(S_{1i}^xe^{i\mathbf{q}_{K}\cdot\mathbf{r}_i}-S_{2i}^xe^{i\mathbf{q}_{K}\cdot\mathbf{r}_i}),\nonumber\\
	\psi^z_{K,\Gamma}&=&\frac{1}{L}\sum_{i}(S_{1i}^ze^{i\mathbf{q}_{K,\Gamma}\cdot\mathbf{r}_i}-S_{2i}^ze^{i\mathbf{q}_{K,\Gamma}\cdot\mathbf{r}_i}),
\end{eqnarray}
where \(\mathbf{q}_K = (\frac{4\pi}{3}, 0)\) and \(\mathbf{q}_\Gamma = (0, 0)\). The minus sign reflects the antiferromagnetic interlayer interaction. \(\psi_K^x\) corresponds to the \(U(1)\) spin rotational symmetry, while \(\psi_K^z\) and \(\psi_\Gamma^z\) are related to the discrete \(\mathbb{Z}_3\) permutation symmetry and \(\mathbb{Z}_2^d\) bilayer dimer symmetry, respectively. We measure the Binder ratio~\cite{Binder1981, Binder2010} \(U = \frac{\langle |\psi|^4 \rangle}{\langle |\psi|^2 \rangle^2}\) 
and the correlation length~\cite{Viet2009,Seabra2011} \(\xi(\mathbf{q}) = \frac{1}{|\delta \mathbf{q}|} \sqrt{\frac{\langle|\psi(\mathbf{q})|^2\rangle}{\langle|\psi(\mathbf{q} + \delta \mathbf{q})|^2\rangle} - 1}\) with \(\delta \mathbf{q} = (2\pi / L, 0)\) for the corresponding order parameters $\psi(\mathbf{q})$. At critical points, the Binder ratio $U$ and correlation length ratio \(\xi / L\) become size-independent, enabling us to determine the critical temperatures \(T_c\).

The phase diagram for \( J_\perp / J = 0.2 \), \( J_\perp^0 / J = 1 \), and \( \Delta = 3 \), shown in Fig.~\ref{fig:Fig4}(a), reveals a rich variety of magnetic phases. Beyond the ``Y'' phase, three additional phases emerge due to \( S_3 \) symmetry breaking: the \( S_3 \)-broken \( tss \) phase, the \(\mathbb{Z}_3\)-broken (\( uud \)) phase, and an intermediate BKT phase. While the exact experimental parameters of \ce{K2Co2(SeO3)2} remain undetermined, the theoretical phase diagram at low fields aligns remarkably well with the experimental results in Fig.~\ref{fig:Fig3}(d), highlighting the validity of the model.

Figure~\ref{fig:Fig4}(b)-(g) illustrate the determination of critical temperatures for a representative magnetic field \( B / J = 0.7 \). From high to low temperatures, the Binder ratio \( U \) for \(\psi_K^z\) [Fig.~\ref{fig:Fig4}(b)] converges at \( T_{c1} = 0.55J \) across different system sizes, signaling a BKT phase transition. This is further supported by the scaled correlation length \(\xi_K^z / L\) [Fig.~\ref{fig:Fig4}(c)], which shows a size-independent behavior in the range \( T_{c2} < T < T_{c1} \) (\( T_{c2} = 0.35J \)), confirming the BKT phase. At \( T_{c3} = 0.135J \), a transition related to the bilayer dimer \(\mathbb{Z}_2^d\) symmetry breaking is observed, as evidenced by the Binder ratio \( U_\Gamma^z \) [Fig.~\ref{fig:Fig4}(d)] and the scaled correlation length \(\xi_\Gamma^z / L\) [Fig.~\ref{fig:Fig4}(e)]. Finally, at \( T_{c4} = 0.051J \), the Binder ratio \( U_K^x \) [Fig.~\ref{fig:Fig4}(f)] and the scaled correlation length \(\xi_K^x / L\) [Fig.~\ref{fig:Fig4}(g)] for \(\psi_K^x\) reveal a BKT-like transition driven by \( U(1) \) symmetry breaking.

\emph{Discussions and Conclusions. --}  
We have investigated the magnetic phases of the bilayer triangular antiferromagnet \ce{K2Co2(SeO3)3}, uncovering a rich interplay of geometric frustration, bilayer coupling, and symmetry-driven phase transitions. The remarkable agreement between experimental  and theoretical phase diagrams in Figs.~\ref{fig:Fig3}(d) and \ref{fig:Fig4}(a) and  at low fields validates the pivotal role of \(S_3 = \mathbb{Z}_3 \otimes \mathbb{Z}_2^d\) in shaping the phase structure. While holographic theory suggests the possibility of a continuous \(S_3 \leftrightarrow \mathbb{Z}_1\) phase transition \cite{Chatterjee2023a}, our results reveal that this transition is not direct but proceeds via two steps, involving an intermediate \(\mathbb{Z}_2\)-symmetric (\(uud\)) phase or a BKT phase. Notably, the BKT phase forms a stable finite-field region in \ce{K2Co2(SeO3)3}, in contrast to the monolayer counterpart~\ce{K2Co(SeO3)2}, where such behavior exists only at zero field. These findings establish \ce{K2Co2(SeO3)3} as an ideal platform for exploring symmetry-driven phase transitions in frustrated magnets and open avenues for studying quantum phenomena and exotic spin dynamics stabilized by bilayer symmetry.

The theoretical phase diagram at low fields is primarily determined by the \(U(1) \otimes S_3\) symmetry in the weak-dimer limit, independent of the specific interaction values. The high-field phases, potentially governed by different symmetries, e.g., for 1/2-plateau phase, have been less thoroughly explored due to uncertainties in the interacting parameters.
By fine-tuning parameters, classical Monte Carlo simulations successfully reproduce the 1/3, 1/2, and 2/3 plateaus (see Fig.S2 in Supplementary Materials~\cite{notes}), but the experimentally observed 5/6 plateau suggests significant quantum effects or different interactions beyond classical models. To address these questions, further neutron scattering experiments may provide insights into the magnetic exchange parameters.
Quantum many-body simulations, such as tensor-network approaches~\cite{Cirac2021}, could provide deeper insights into the role of bilayer coupling in amplifying frustration and stabilizing unconventional magnetic states, such as the 5/6-plateau phase. These methods would also enable a more comprehensive exploration of the influence of symmetry in governing quantum phase transitions at zero temperature.

\acknowledgments
This work is supported by the National Natural Science Foundation of China (Grant No.~12474143, 12404172, and 12374129), the National Key Research and Development Program of China (Grant No.~2021YFA1400400), Shenzhen Fundamental Research Program (Grant No.~JCYJ20220818100405013  and JCYJ20230807093204010), Anhui Provincial Natural Science Foundation No. 2408085J025 and the Anhui Provincial Major S \& T Project No. s202305a12020005. A portion of this work was carried out at the Steady High Magnetic Field Facilities in High magnetic Field Laboratory (CAS), Synergetic Extreme Condition User Facility (SECUF) and Wuhan National High Magnetic Field Center.

\bibliography{refs}

\end{document}